\RequirePackage{fix-cm}
\documentclass[smallextended]{svjour3}       % onecolumn (second format)
\smartqed  % flush right qed marks, e.g. at end of proof
\usepackage{graphicx}
\usepackage{txfonts}
\usepackage{graphicx}
\usepackage{txfonts}
\usepackage{array}
\usepackage{color}
\usepackage{url}
\usepackage{enumitem}
\usepackage{bm}
\usepackage{xcolor}
\usepackage{aas_macros}
\usepackage[colorlinks = true,
            linkcolor = blue,
            urlcolor  = blue,
            citecolor = blue,
            anchorcolor = blue]{hyperref}
\journalname{Experimental Astronomy}
\begin{document}

\title{The Ariel 0.6 - 7.8 $\mu$m stellar limb-darkening coefficients %\thanks{Grants or other notes
%about the article that should go on the front page should be
%placed here. General acknowledgments should be placed at the end of the article.}
}
%\subtitle{Subtitle}

%\titlerunning{Short form of title}        % if too long for running head

\author{Giuseppe Morello         \and
        Camilla Danielski \and Subhajit Sarkar%etc.
}

%\authorrunning{Short form of author list} % if too long for running head

\institute{Morello G. \at
              AIM, CEA, CNRS, Universit\'e Paris-Saclay, Universit\'e Paris Diderot, Sorbonne Paris Cit\'e, \\ 
              F-91191 Gif-sur-Yvette, France \\
              %Tel.: \red{XXX}\\
              %Fax:\red{XXX}\\
              %\at
              \emph{Present address:} Instituto de Astrof\'isica de Canarias (IAC), 38200, La Laguna, Tenerife, Spain \\
              \email{gmorello@iac.es}  \\
              ORCID: 0000-0002-4262-5661   %  \\
%             \emph{Present address:} of F. Author  %  if needed
           \and
            Danielski C.\at
            UCL Centre for Space Exochemistry Data \\
            Atlas Building,  Fermi Avenue, Harwell Campus,  Didcot,  OX11 0QR (UK)\\
            \emph{Present address:} Instituto de Astrof\'isica de Andaluc\'ia (IAA-CSIC)\\
            Glorieta de la Astronom\'ia s/n, 18008 Granada, Spain\\
            ORCID: 0000-0002-3729-2663
            \and
           Sarkar S.\at
           School of Physics and Astronomy, Cardiff University,\\ The Parade, Cardiff, CF24 3AA, UK
}

\date{Received: date / Accepted: date}
% The correct dates will be entered by the editor

\maketitle

%\noindent Link with guidelines is \url{https://www.springer.com/journal/10686/submission-guidelines}

\begin{abstract}
We provide here tables of stellar limb-darkening coefficients (LDCs) for the Ariel ESA M4 space mission. These tables include LDCs corresponding to different wavelength bins and white bands for the NIRSpec, AIRS-Ch0 and AIRS-Ch1 spectrographs, and those corresponding to the VISPhot, FGS1 and FGS2 photometers.
The LDCs are calculated with the open-source software \texttt{ExoTETHyS} for three complete grids of stellar atmosphere models obtained with the \texttt{ATLAS9} and \texttt{PHOENIX} codes. The three model grids are complementary, as the \texttt{PHOENIX} code adopts more modern input physics and spherical geometry, while the models calculated with \texttt{ATLAS9} cover wider ranges of stellar parameters. The LDCs obtained from corresponding models in the \texttt{ATLAS9} and \texttt{PHOENIX} grids are compared in the main text. All together the models cover the following ranges in effective temperature ($1\,500 \, K \le T_{\mbox{\small eff}} \le 50\,000 \, K$), surface gravity (0.0 dex $\le \log{g} \le 6.0$ dex), and metallicity ($-5.0 \le [M/H] \le 1.0$).

\end{abstract}

\section{Introduction}
\label{sec:intro}
Observations of exoplanetary transits are affected by the specific intensity decrease from the centre to the edge of the projected stellar disc, so-called limb-darkening effect \cite{mandel2002}. While the planet's projection moves onto the stellar disc, the fraction of occulted stellar flux varies over time, instead of being equal to the planet-to-star area ratio.
Furthermore, the limb-darkening is a wavelength-dependent effect, meaning that it is crucial to account for it when modelling transit light-curves at multiple frequencies
\cite{knutson2007,morello2017,morello2018,danielski2020}. A precise determination of the limb-darkening will consequently enable a precise characterisation of an exoplanet's atmosphere.

One of the Ariel mission requirements is to meet a photometric precision of 10-50 parts per million (ppm) relative to the stellar flux of the observed targets \cite{tinetti2018,edwards2019}. 
Here we provide grids of limb-darkening coefficients (LDCs) for the Ariel instruments, adopting the four-coefficient law introduced by Claret (2000) \cite{claret2000}, hereinafter referred to as ``claret-4''. In particular, we used the software \texttt{ExoTETHyS} \cite{morello2020,morello2020joss} to compute LDCs with the there-called \texttt{ATLAS\_2000} \cite{claret2000}, \texttt{PHOENIX\_2012\_13} and \texttt{PHOENIX\_DRIFT\_2012} \cite{claret2012,claret2013} model grids.
The spectral responses of the Ariel instruments were obtained through ExoSim \cite{sarkar2020}. We considered wavelength bins of 0.05 $\mu$m for the NIRSpec and AIRS-Ch1 spectrometers, and 0.02 $\mu$m for AIRS-Ch0.
The tables attached to this manuscript contain the most up to date LDCs which can be used by the Ariel community during the preparatory studies. We note that, the underlying stellar and instrumental models are likely to be updated by the launch of the Ariel spacecraft. Finally, whether a specific set of LDCs is not available here (e.g., adopting different limb-darkening laws, wavelength bins, or to interpolate between the stellar parameters of the grid), it can be computed with the python package \texttt{ExoTETHyS}\footnote{Main GitHub repository \url{https://github.com/ucl-exoplanets/ExoTETHyS}}.

The manuscript is structured as follows. Section \ref{sec:ld}  introduces the stellar limb-darkening effect, and discusses its historical overview, putting particular emphasis on its application to the study of exoplanetary transits. Section \ref{sec:exotethys} briefly describes the \texttt{ExoTETHyS} limb-darkening subpackage used to calculate the LDCs for the Ariel mission. Section \ref{sec:ariel} provides information about the Ariel science instruments. Section \ref{sec:conclusions} summarises this paper's content.

\section{The stellar limb-darkening effect}
\label{sec:ld} 

\subsection{Limb-darkening laws}
\label{sec:ldlaws}
The phenomenon of limb-darkening is easily understood starting from the formal solution of the radiative transfer equation along a ray, which shows that the emerging intensity is a weighted average of the source function over optical depth. 
Historically, the approximation that was usually adopted is the Eddington-Barbier one, which describes the source function of the continuum radiation as a polynomial of optical depth in a medium with plane-parallel geometry. In this case, the intensity emerging at an angle $\theta$ with respect to the surface normal is equal to the source function at optical depth $\mu = \cos{\theta}$. The first parametrization proposed to model stellar limb-darkening is the linear law \cite{schwarzschild1906}:
\begin{equation}
\label{eqn:ld_law_linear}
\frac{I_{\lambda}(\mu)}{I_{\lambda}(1)} = 1 - a(1-\mu) ,
\end{equation}
as derived from the first-order Eddington-Barbier approximation.
It was later realised that the linear law does not fully capture the centre-to-limb intensity variation, thus a quadratic law was preferred \cite{kopal1950}:
\begin{equation}
\label{eqn:ld_law_quadratic}
\frac{I_{\lambda}(\mu)}{I_{\lambda}(1)} = 1 - a_1(1-\mu) - a_2(1-\mu)^2 .
\end{equation}
Later, many researchers have proposed a long list of other limb-darkening laws, which are not only polynomial functions of $\mu$ (see \cite{kipping2013,espinoza2016} for reviews). We highlight here the power-2 law \cite{hestroffer1997}:
\begin{equation}
\label{eqn:ld_law_power2}
\frac{I_{\lambda}(\mu)}{I_{\lambda}(1)} = 1 - c(1-\mu^{\alpha}),
\end{equation}
which outperforms other biparametric laws, especially for the M-dwarf models \cite{morello2017,maxted2018}. The search for the simplest law that accurately describes the stellar intensity profile was motivated by both the aim of linking the LDCs to physical parameters, and the fact that fewer coefficients are more easily manageable as free parameters. However, no biparametric law turns out to be sufficiently accurate for all stellar types and wavelengths at the level of precision in the data achieved with modern instruments \cite{espinoza2016,morello2017}. For this reason we strongly recommend the use of the claret-4 law \cite{claret2000}:
\begin{equation}
\label{eqn:ld_law_claret4}
\frac{I_{\lambda}(\mu)}{I_{\lambda}(1)} = 1 - \sum_{k=1}^{4} a_k(1-\mu^{k/2}) ,
\end{equation}
since it has proven highly reliable in all the investigations carried out so far \cite{howarth2011b,magic2015,claret2020,morello2020}.

The above limb-darkening laws may not reproduce the behaviour of some spectral lines at high-resolution ($R>10^4$), whereas more complex intensity profiles, including limb-brightening effects, can occur \cite{czesla2015}. The discussion of these cases is not relevant to the Ariel instruments, which provide optical broadband photometry and infrared low-resolution spectroscopy with $R \sim 15\mbox{-}100$.

\begin{figure}
    \centering
    \includegraphics[width=\textwidth]{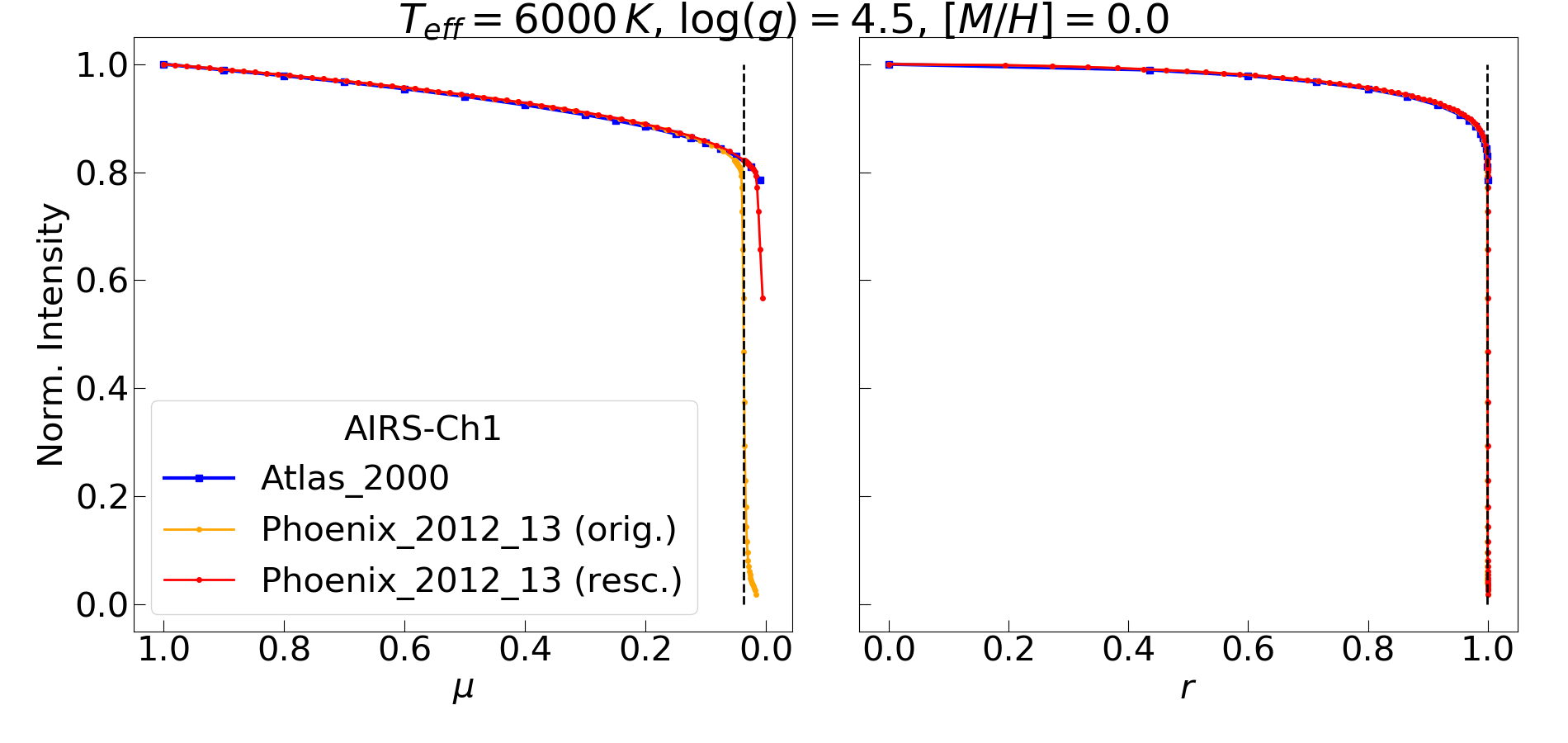}
    \caption{Left panel: Normalised intensities vs. $\mu$ for a solar-like star integrated over the AIRS-Ch1 white band. Note the intensity drop-off obtained with the ``original'' \texttt{PHOENIX} models (orange) due to the spherical geometry, then rescaled (red). The \texttt{ATLAS} models (blue) assume plane-parallel geometry. The vertical dashed line indicates the point where the \texttt{PHOENIX} models are truncated in the LDCs fit with \texttt{ExoTETHyS}. Right panel: Analogous plot with normalised intensities vs. $r$.}
    \label{fig:pp_sph_compar}
\end{figure}

\subsection{Plane-parallel versus spherical geometry}
\label{sec:sphericalLD}
The earliest models of stellar atmospheres assumed plane-parallel geometry (e.g., \cite{kurucz1979,wade1985,diaz-cordoves1995,claret1995,claret2000,claret2004,claret2008,sing2010,howarth2011b,reeve2016}). 
Recently, several models that take into account the spherical geometry of the stellar surface have appeared in the literature (e.g., \cite{claret2011,claret2012,claret2013,neilson2013a,neilson2013b,claret2014,magic2015,claret2018,claret2020}).
The intensity profiles obtained with spherical geometry show a characteristic steep drop-off, such that the intensities are almost zero at small but finite $\mu$ values (see Figure \ref{fig:pp_sph_compar}). We note that the intensity drop-off in the infrared is more abrupt than the one in the UV and visible.
This behaviour is well understood \cite{wittkowski2004,espinoza2015,morello2017}.
The rays connecting the observer with the star surface at small $\mu$ values intersect only the outermost atmospheric shells, which have negligible emissivity compared to more internal shells.
Instead, any optical ray intersects all the layers of a plane-parallel atmosphere, so that the emerging intensity always includes the contributions of inner layers down to optical depth of unity. Consequently, the intensity profile appears to be a smooth function of $\mu$ with a non-zero limit at $\mu=0$.

Limb-darkening laws such as those of Equations \ref{eqn:ld_law_linear}-\ref{eqn:ld_law_claret4} were designed to approximate the stellar intensity profiles obtained with plane-parallel geometry, but they cannot reproduce the steep drop-off resulting from spherical geometry.
Some authors have suggested applying a cut-off in $\mu$, i.e. discarding the points with $\mu < \mu_0$ when fitting the intensity profile with a limb-darkening law \cite{claret2003,claret2012,morello2020}. The truncated profiles are referred to as quasi-spherical models. Claret (2018) \cite{claret2018} replaced the after drop-off intensities with zeroes, instead of discarding them.
Also, the drop-off radius corresponds to what would appear as the stellar radius in observations of exoplanetary transits, interferometric imaging, or lunar occultation. Therefore, the spherical intensity profiles should be rescaled so that $\mu = 0$ at the inflection point of the intensity profile \cite{wittkowski2004,espinoza2015,morello2017}.

\subsection{Optimisation fitting methods}
\label{sec:optimisation}
The optimisation method plays an important role when fitting for LDCs.
The simplest approach is the standard least-squares fit to the precalculated intensities. The least-squares method is sensitive to the choice of sampling of the intensity profile. The cause of sampling dependence is the impossibility to obtain a perfect match to the reference profile with the chosen parametrization, especially when using one or two-coefficient laws, and/or considering spherical geometry. A uniform sampling in $\mu$ gives more weight to the regions on the edge of the stellar disc. In fact, the radial distance of a $\mu$-point from the center of the disc is
\begin{equation}
\label{eqn:rmu}
r= \sqrt{1 - \mu^2} ,
\end{equation}
where $r=1$ ($\mu=0$) denotes the contour of the disc. In the case of uniform sampling in $\mu$, $50\%$ of the weight is attributed to the annulus with $r > 0.866$. Typically the situation is even more unbalanced, given that the spherical models adopt a finer sampling near the intensity drop-off. In contrast, a uniform sampling in $r$ tends to undersample the region where the effect of limb-darkening is more pronounced.
The matter of optimal sampling to calculate the LDCs has been widely debated in the previous literature \cite{heyrovsky2007,claret2008,claret2011,parviainen2015}.

Flux-conservation has been considered as an additional condition or alternative to least-squares fitting \cite{wade1985,claret2000,howarth2011}. It requires that the integral of the flux over the disc obtained with the limb-darkening law is exactly equal to the stellar flux of the model, a requirement that is not automatically satisfied by the best-fit least-squares solution.

%REMOVED AFTER REFEREE1'S COMMENT We refer to section \ref{sec:sphericalLD} for a discussion about the treatment of the intensity drop-off in stellar models with spherical geometry.

\subsection{The Synthetic-Photometry/Atmosphere-Model LDCs}
Howarth (2011) \cite{howarth2011} raised the question of which is the best set of LDCs to be adopted in the light-curve analyses of exoplanetary transits. He generated synthetic-photometry/atmosphere-model (SPAM) light-curves for well-studied systems, using as inputs the empirically determined geometric parameters, and the intensity distribution corresponding to the known stellar parameters (approximated by a set of claret-4 LDCs, Equation \ref{eqn:ld_law_claret4}). Light-curve fitting was then performed with one or two LDCs as free parameters (e.g., Equations \ref{eqn:ld_law_linear}-\ref{eqn:ld_law_quadratic}). The resulting SPAM-LDCs were significantly different from those obtained by fitting on the intensity profile with various optimisation methods.

The follow-up work by Morello et al. (2017) \cite{morello2017} on this study used full spherical intensity models with numerical integration to generate synthetic light-curves, and different configurations for light-curve fitting. They noted that the intensity profiles inferred by light-curve fitting reproduce the behaviour of the models well inside the disc, but often deviate at the smaller $\mu$ values. 
The corresponding transit depths and orbital parameters can also be significantly biased. The apparent good quality of the light-curve fitting can be deceptive.

Part of the problem is due to the inadequacy of a given limb-darkening law to approximate the model intensity distribution on the whole stellar disc. Therefore, different optimisation methods, including SPAM optimisation, attribute different weights to the regions of the disc. The residuals of the light-curve fits do not necessarily provide a useful diagnostic, because the strong degeneracies between LDCs and geometric parameters can lead to an excellent match to the transit light-curves, even with the wrong parameter values.

All these historical issues have been addressed in the  \texttt{ExoTETHyS} package \cite{morello2020} which we briefly describe in the upcoming section.

\begin{table}[]
    \centering
    \begin{tabular}{c|c|c|c|c|c}
        %Name & Geometry & ${\rm \Delta} T_{\mbox{\small eff}}$ ($K$) & ${\rm \Delta} \log{g}$ (dex) & ${\rm \Delta} [M/H]$ & ${\rm \Delta} \lambda$ ($\mu$m) \\
        %Name & Geometry & Range $T_{\mbox{\small eff}}$ ($K$) & Range $\log{g}$ (dex) & Range $[M/H]$ & Range $\lambda$ ($\mu$m) \\
        Name & Geometry & $T_{\mbox{\small eff}}$ ($K$) & $\log{g}$ (dex) & $[M/H]$ & $\lambda$ ($\mu$m) \\
        \hline
        \texttt{ATLAS\_2000} & P-P & 3500-50000 & 0.0-5.0 & --5.0-1.0 & 0.009-160.0 \\
        \texttt{PHOENIX\_2012\_13} & S1 & 3000-10000 & 0.0-6.0 & 0.0 & 0.25-10.0 \\
        \texttt{PHOENIX\_DRIFT\_2012} & S1 & 1500-3000 & 2.5-5.5 & 0.0 & 0.001-950.0 \\
    \end{tabular}
    \caption{Information about the stellar model grids adopted in this work. P-P = plane-parallel, S1 = spherical 1D geometry.}
    \label{tab:grids}
\end{table}

\section{The ExoTETHyS package}
\label{sec:exotethys}
We state here the conditions that must be fulfilled by an optimal set of LDCs for a mission dedicated to the characterisation of transiting exoplanets:
\begin{enumerate}
    \item if the optimal LDCs are kept fixed during light-curve fitting, the best-fit transit depth and orbital parameters should be unbiased;
    \item the LDCs can be computed directly from stellar intensity profiles, e.g., without the need for lengthy calculations involving \emph{ad hoc} synthetic light-curves.
\end{enumerate}
The software \texttt{ExoTETHyS.SAIL} is a calculator of LDCs that has been designed to meet the above requirements \cite{morello2020joss}. It performs a weighted least-squares fit to the normalised stellar intensities ($I_{\lambda}(\mu)/I_{\lambda}(1)$). The weigths are proportional to the sampling interval in $r$, with a cut-off of $r \le 0.99623$ for the intensity profiles obtained with spherical geometry, after rescaling (see Section \ref{sec:sphericalLD}). Flux-conservation poses no additional constraints when fitting normalised intensities, since the absolute flux can be matched by adjusting the multiplicative factor (approximately equal to $I_{\lambda}(1)$). The absolute flux value does not affect the transit light-curve morphology and the relevant astrophysical parameters.

In particular, the LDCs calculated according to the claret-4 law (Equation \ref{eqn:ld_law_claret4}) have been tested to guarantee a precision of $<$10 ppm over the whole light-curve model at all wavelengths \cite{morello2020}.
This level of precision is up to 2 orders of magnitude better than that obtained with other existing software that use different optimisation algorithms. It is worth noting that the improvement of \texttt{ExoTETHyS.SAIL} over other algorithms is particularly significant in the infrared, given its capability of better dealing with the more abrupt intensity drop-off (see Section \ref{sec:sphericalLD}).

\paragraph{Stellar atmosphere models}~\\
\vspace{0.5mm}\\
At the time of writing, the \texttt{ExoTETHyS} package contains 3 databases with pre-computed stellar model intensities that fully cover the wavelength range of the Ariel instruments (see Table \ref{tab:grids}).

The \texttt{ATLAS\_2000} database includes the largest variety of stellar types, covering wide ranges in effective temperature, surface gravity, and metallicity. The corresponding models of stellar atmospheres were calculated by Kurucz (1993, 1996) \cite{kurucz1993,kurucz1996} using the \texttt{ATLAS9} software. The intensity spectra produced with these models were previously used by Claret (2000) \cite{claret2000}. We selected those models with microturbulent velocity of $2 \, km/s$. Some other models with different values of the microturbulent velocity were available for stars with solar metallicity, but they were not included in the \texttt{ExoTETHyS} database as the variations of the LDCs with this parameter are negligible.

The \texttt{PHOENIX\_2012\_13} database focuses on stars of later types, such as the typical exoplanet host stars, and with solar metallicity. The corresponding models of stellar atmospheres were generated with the \texttt{PHOENIX} code, assuming phase equilibrium (\texttt{DUSTY/COND} physics). 

The \texttt{PHOENIX\_DRIFT\_2012} database is specific to the coolest stars down to $1\,500 \, K$. The corresponding models of stellar atmospheres were generated with the \texttt{PHOENIX} code, including a detailed computation of high-temperature condensate clouds (\texttt{DRIFT} physics) \cite{witte2009}.

The two \texttt{PHOENIX} databases were produced expressly for computing LDC tables \cite{claret2012,claret2013}.

For targets whose LDCs can be obtained from both \texttt{ATLAS} and \texttt{PHOENIX} tables, we recommend using the \texttt{PHOENIX} values, as they rely on newer input physics and chemistry, and more realistic geometry. For stars with $T_{\mbox{\small eff}} = 3\,000 \, K$ , the \texttt{DRIFT} models are more accurate than \texttt{DUSTY/COND}.
We refer to the relevant literature for the comparison between the \texttt{ATLAS} and \texttt{PHOENIX} algorithms \cite{kurucz1979,allard1995,allard2001,castelli2004,witte2009,husser2013}. Section \ref{sec:ariel} adds the comparison between intensity profiles obtained in the Ariel passbands.

New grids of stellar atmosphere models with updated input physics and chemistry are being prepared and will be available for the upcoming phases of the Ariel mission.
The new information will be incorporated into the \texttt{ExoTETHyS} package.

\section{LDCs for the Ariel mission}
\label{sec:ariel}

\begin{table}[]
    \centering
    \begin{tabular}{c|c|c}
        Name &  ${\rm \Delta} \lambda$ ($\mu$m) & Spectral Resolving Power \\
        \hline
        VISPhot & 0.5-0.6 & Integrated band \\
        FGS1 & 0.6-0.8 & Integrated band \\
        FGS2 & 0.8-1.1 & Integrated band \\
        NIRSpec & 1.1-1.95 & $R \ge$15 \\
        AIRS-Ch0 & 1.95-3.9 & $R \ge$100 \\
        AIRS-Ch1 & 3.9-7.8 & $R \ge$30 \\
    \end{tabular}
    \caption{Nominal wavelength ranges and required binned spectral resolving power for the 6 Ariel science instruments (Ariel Payload Requirements Document, ARIEL-RAL-PL-RS-001, Issue 3.0, 31 July 2020).}
    \label{tab:instruments}
\end{table}

\begin{figure}
    \centering
    \includegraphics{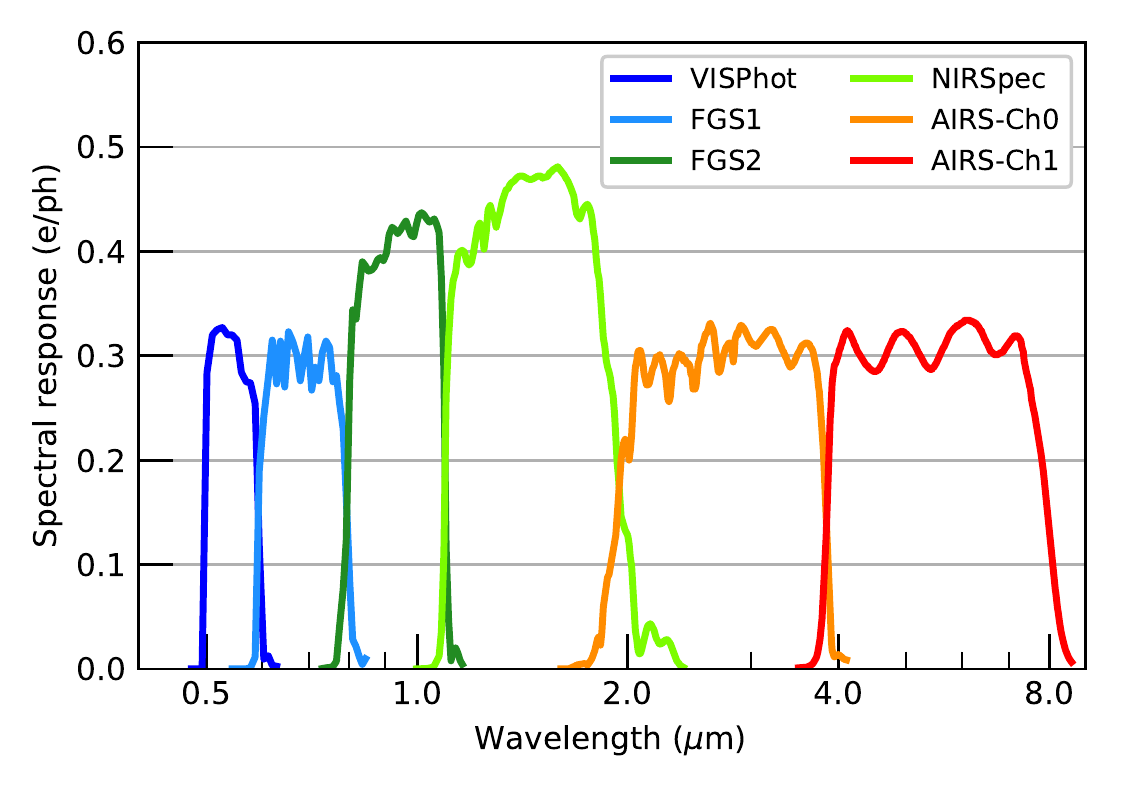}
    \caption{The calculated photon-to-electron conversion efficiencies (PCEs) for the 6 Ariel science instruments.}
    \label{fig:PCEs}
\end{figure}

The Ariel instruments include 3 broadband photometers at 0.5-1.1 $\mu$m, and 3 infrared spectrometers covering the 1.1-7.8 $\mu$m wavelength range. Table \ref{tab:instruments} reports basic information on these instruments.
The known spectral response (Figure \ref{fig:PCEs}) of the Ariel instruments has been taken into account for the calculation of the LDCs. The response for each instrument was obtained by multiplying the detector quantum efficiency and the total transmission of the optical train, as obtained through ExoSim \cite{sarkar2020}. 

We provide tables of LDCs for all the instruments of the Ariel mission. These tables have been pre-calculated for the users convenience and can be found in the supplementary material provided online.
Table \ref{tab:ldc_example} shows an example with the content of a file  (e.g., name of instrument, stellar model, wavelength range, etc.).
We performed calculations on wavelength bins of 0.05 $\mu$m for the NIRSpec and AIRS-Ch1 spectrometers, and 0.02 $\mu$m for AIRS-Ch0. These wavelength bins are narrower than or equal to those corresponding to the required binned spectral resolving power of the instruments (see Table \ref{tab:instruments}). This information is subject to small changes during the detailed definition phase C.

\begin{table}[]
\texttt{PHOENIX\_2012\_13} \hspace{1cm} \texttt{AIRS\_ch0}\\[0.1cm]
\resizebox{\textwidth}{!}{%
    \centering
    \begin{tabular}{c|c|c|c|c|c|c|c|c}
        $T_{\mbox{\small eff}}$ ($K$) & $\log{g}$ (dex) & $[M/H]$ (dex) & $\lambda_1$ ($\mu$m) & $\lambda_2$ ($\mu$m) & $a_1$ & $a_2$ & $a_3$ & $a_4$ \\
        \hline
        3000.0 & 1.50 & 0.00 & 1.940 & 3.900 & 4.77677453 & -7.89734570 & 6.82914503 & -2.19767201 \\
        3000.0 & 1.50 & 0.00 & 1.940 & 1.960 & 2.50560261 & -2.90202256 & 2.56740110 & -0.87192810 \\
        3000.0 & 1.50 & 0.00 & 1.960 & 1.980 & 2.85438826 & -3.73580417 & 3.32761458 & -1.10490417 \\
        3000.0 & 1.50 & 0.00 & 1.980 & 2.000 & 2.20836574 & -2.40865397 & 2.16929277 & -0.73787659 \\
        $\cdots$ & $\cdots$ & $\cdots$ & $\cdots$ & $\cdots$ & $\cdots$ &  $\cdots$ & $\cdots$ & $\cdots$ \\
       3000.0 & 1.50 & 0.00 & 3.880 & 3.900 & 5.91672785 & -10.51632605 & 9.03420722 & -2.90162989 \\
       3000.0 & 2.00 & 0.00 & 1.940 & 3.900 & 1.21751542 & -0.93156054 & 0.80975259 & -0.27133061 \\
       3000.0 & 2.00 & 0.00 & 1.940 & 1.960 & 0.20321759 & 1.38089364 & -1.02166342 & 0.26353188 \\
       $\cdots$ & $\cdots$ & $\cdots$ & $\cdots$ & $\cdots$ & $\cdots$ &  $\cdots$ & $\cdots$ & $\cdots$ \\
    \end{tabular}
    }
    \caption{Example of LDCs table extracted from supplementary file \texttt{ARIEL\_grid\_ldc\_phoenix\_cond\_airsch0.txt}. The first line contains the stellar model database and the Ariel instrument name. The second line defines the column entries of the following lines: stellar effective temperature, surface gravity, metallicity, wavelength intervals, and claret-4 LDCs, respectively.}
    \label{tab:ldc_example}
\end{table}

\begin{figure}
    \centering
    \includegraphics[width=\textwidth]{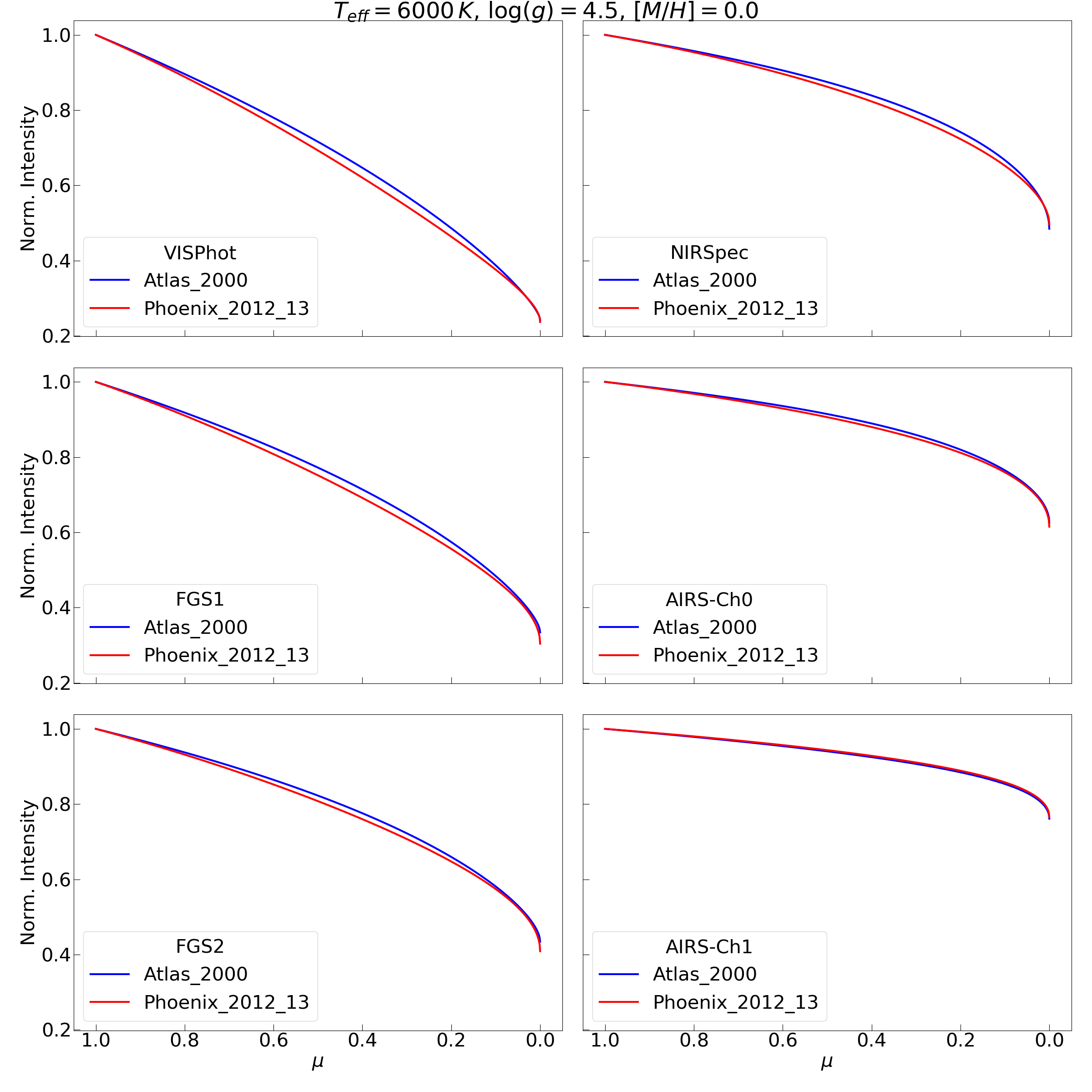}
    \caption{Limb-darkening profiles for a solar-like star using the LDCs from the \texttt{ATLAS\_2000} (blue) and \texttt{PHOENIX\_2012\_13} (red) tables provided in the supplemental material. The left panels show the results obtained for the VISPhot, FGS1 and FGS2 photometers. The right panels show the results for the white band integrated spectrometers.}
    \label{fig:grid_compar_solar}
\end{figure}

\begin{figure}
    \centering
    \includegraphics[width=\textwidth]{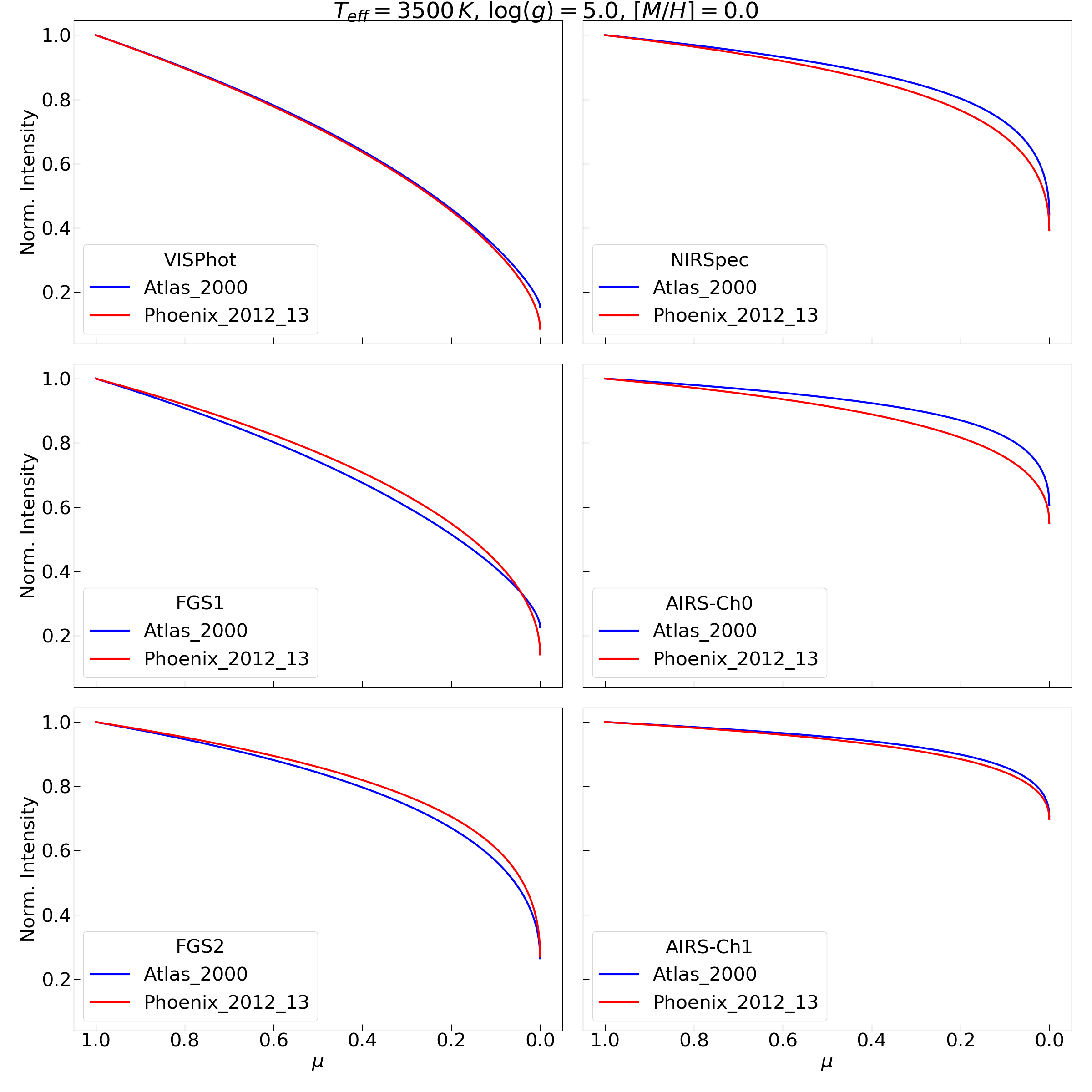}
    \caption{Analogous to Figure \ref{fig:grid_compar_solar}, but for an M dwarf.}
    \label{fig:grid_compar_mdwarf}
\end{figure}

Figures \ref{fig:grid_compar_solar} and \ref{fig:grid_compar_mdwarf} compare the limb-darkening profiles calculated with the \texttt{ATLAS\_2000} and \texttt{PHOENIX\_2012\_13} databases for a solar-like star and a M dwarf, respectively. The two sets of profiles for the solar-like star are in good agreement for the 6 Ariel instruments, although the discrepancy in the visible photometry might be significant at $>$10 ppm level. The M dwarf profiles show greater disagreement, most likely due to the enhanced effect of molecular absorption. The comparison between all the analogous models from the two grids is beyond the scope of this paper.

\section{Conclusions}
\label{sec:conclusions}
We discussed here the stellar limb-darkening effect on the study of transiting exoplanets and their atmospheres. We documented an open-source software, part of the \texttt{ExoTETHyS} package, which has been specifically designed to achieve a precision of a few parts per million in modelling the transit light-curves. By using \texttt{ExoTETHyS} and the  
spectral responses of the Ariel instruments, obtained through ExoSim, we provide pre-calculated tables of limb-darkening coefficients over 0.6-7.8 $\mu$m.
The software is also available to calculate limb-darkening coefficients with settings personalised by the user, e.g., different spectral bins. The current calculations rely on \texttt{ATLAS9} and \texttt{PHOENIX} stellar intensity spectra, which cover an overall effective temperature range of 1500-50000 $K$, surface gravity range of 0.0-6.0 dex, and metallicity of --5.0-1.0 dex. New grids of stellar atmosphere models covering the Ariel wavelength range of 0.6-7.8 $\mu$m are being prepared for inclusion in \texttt{ExoTETHyS}.

\begin{acknowledgements}
The authors would like to thank A. Claret and E. Pascale for useful discussions. G.M. was supported by the LabEx P2IO, the French ANR contract 05-BLANNT09-573739. S.S. was supported by United Kingdom Space Agency (UKSA) grant: ST/S002456/1.
%If you'd like to thank anyone, place your comments here
%and remove the percent signs.
\end{acknowledgements}

% BibTeX users please use one of
%\bibliographystyle{spbasic}      % basic style, author-year citations
\bibliographystyle{spmpsci}      % mathematics and physical sciences
\bibliography{MAIN.bib}   % name your BibTeX data base

% Non-BibTeX users please use
%\begin{thebibliography}{}
%
% and use \bibitem to create references. Consult the Instructions
% for authors for reference list style.
%
%\bibitem{RefJ}
% Format for Journal Reference
%Author, Article title, Journal, Volume, page numbers (year)
% Format for books
%\bibitem{RefB}
%Author, Book title, page numbers. Publisher, place (year)
% etc
%\end{thebibliography}

\end{document}